\documentclass[a4paper]{jpconf}
\usepackage{graphicx}
\usepackage{siunitx}
\begin{document}

\title{Results from the Project 8 phase-1 cyclotron radiation emission spectroscopy detector}

\author{
        A~Ashtari~Esfahani$^1$,
        S~B{\"o}ser$^2$,
        C~Claessens$^2$,
        L~de~Viveiros$^3$,
        P~J~Doe$^1$,
        S~Doeleman$^4$,
        M~Fertl$^1$,
        E~C~Finn$^5$,
        J~A~Formaggio$^6$,
        M~Guigue$^5$,
        K~M~Heeger$^7$,
        A~M~Jones$^5$,
        K~Kazkaz$^8$,
        B~H~LaRoque$^3$,
        E~Machado$^1$,
        B~Monreal$^3$,
        J~A~Nikkel$^7$,
        N~S~Oblath$^5$,
        R~G~H~Robertson$^1$,
        L~J~Rosenberg$^1$,
        G~Rybka$^1$,
        L~Salda{\~n}a$^7$,
        P~L~Slocum$^7$,
        J~R~Tedeschi$^5$,
        T~Th{\"u}mmler$^9$,
        B~A~Vandevender$^5$,
        M~Wachtendonk$^1$,
        J~Weintroub$^4$,
        A~Young$^4$ and
        E~Zayas$^6$
       }

\address{
         $^1$ Center for Experimental Nuclear Physics and Astrophysics, and Department of Physics, University of Washington, Seattle, WA, USA\\
         $^2$ Johannes Guttenberg University, Mainz, Germany\\
         $^3$ Department of Physics, University of California, Santa Barbara, CA, USA\\
         $^4$ Harvard-Smithsonian Center for Astrophysics, Cambridge, MA, USA\\
         $^5$ Pacific Northwest National Laboratory, Richland, WA, USA\\
         $^6$ Laboratory for Nuclear Science, Massachusetts Institute of Technology, Cambridge, MA, USA\\
         $^7$ Department of Physics, Yale University, New Haven, CT, USA\\
         $^8$ Lawrence Livermore National Laboratory, Livermore, CA, USA\\
         $^9$ Karlsruhe Institute for Technology, Karlsruhe, Germany
        }
\ead{laroque@physics.ucsb.edu}

\begin{abstract}
The Project 8 collaboration seeks to measure the absolute neutrino mass scale by means of precision spectroscopy of the beta decay of tritium.
Our technique, cyclotron radiation emission spectroscopy, measures the frequency of the radiation emitted by electrons produced by decays in an ambient magnetic field.
Because the cyclotron frequency is inversely proportional to the electron's Lorentz factor, this is also a measurement of the electron's energy.
In order to demonstrate the viability of this technique, we have assembled and successfully operated a prototype system, which uses a rectangular waveguide to collect the cyclotron radiation from internal conversion electrons emitted from a gaseous $^{83m}$Kr source.
Here we present the main design aspects of the first phase prototype, which was operated during parts of 2014 and 2015.
We will also discuss the procedures used to analyze these data, along with the features which have been observed and the performance achieved to date.
\end{abstract}

\section{Introduction}
The absolute scale of the neutrino masses is a significant gap in our understanding of fundamental particle physics.
While it is possible to derive stringent limits from cosmological observations or searches for neutrinoless double-beta decay, these limits depend implicitly on the models used in their determination.
On the other hand, kinematically derived limits are agnostic to the underlying physics and therefore offer a complementary approach.
The Project 8 collaboration will use a novel technique known as Cyclotron Radiation Emission Spectroscopy (CRES), which was first proposed in 2009 \cite{p8prd}, to make such a direct measurement through spectroscopy of tritium beta decay.

Spectroscopy of tritium beta decay is well established as a technique for such direct measurements, providing both the current best limits from the Mainz and Troitsk experiments,\cite{mainz, troitsk} and KATRIN, the current generation experiment \cite{katrin}.
While established technologies use electrostatic barrier potentials and counting detectors to measure the energy spectrum, the CRES technique leverages the fact that there are relativistic corrections to a particle's cyclotron frequency, and therefore a frequency measurement can be used to determine a particle's total energy.
CRES therefore provides a complementary approach to traditional techniques, subject to different systematic effects and sensitivity scaling constraints.

The Project 8 collaboration constructed the phase 1 detector with the primary goal of demonstrating the CRES technique.
The prototype uses a section of waveguide, sealed with kapton foil windows, to contain $^{83m}$Kr gas, a source of internal-conversion electrons.
The emitted cyclotron radiation is colleted by waveguide and propagates to a receiver.
Coils wrapped around the waveguide provide magnetic field perturbations which confine electron motion parallel to the local magnetic field, while the cyclotron motion itself confines perpendicular motion.
The entire waveguide assembly is inserted into an NMR magnet which provides an axial magnetic field of \SI{0.9467}{T}.
We initially operated the detector during 2014, with the first results published in early 2015 \cite{p8prl}.
Given the success of phase 1, the prototype has been used to study the performance of the technique as an R\&D effort prior to phase 2, where CRES will be applied to the tritium spectrum for the first time.
Here we discuss progress made with the phase 1 detector since publication of our first results.

\section{Hardware changes and improvements}
The essential features of the phase 1 detector systems were described in the publication of first results \cite{p8prl}.
Once data collection was completed in 2014, a comprehensive survey of the analog signal path was conducted.
Several components in the first analog mixing stage, which downconverts signals which start around \SI{26}{GHz} by \SI{24.2}{GHz}, were upgraded.
Additionally, standard maintenance procedures to the cryocooler resulted in a significant improvement to thermal stability of the cryogenic system, including the low noise amplifiers.

For the initial data collection, both a continuously streaming digitizer and a real time spectrum analyzer, with the ability to trigger on electron events, were used to collect data.
The streaming acquisition system can be used to study triggering systematics since it is immune to them, at the cost of producing a much large volume of data; however, the system had to be decommissioned as a result of unexpected hardware failures.
All data taken since our initial publication has been conducted using the triggered data acquisition system exclusively.
A new acquisition system with support for both streaming and triggered modes will be commissioned for phase 2.
Studies of trigger efficiency and investigations related to event rate will be done using that system.

Finally, the initial data collected in the phase 1 system were all taken using a single trapping coil, which generated a local region of slightly reduced magnetic field magnitude.
A careful survey of the magnetic properties of all materials near the trapping region was conducted and several magnetic components were identified and relocated or replaced with non-magnetic alternatives.
As a result, we were able to achieve axial confinement between two coils, each producing a locally increased magnetic field.
This configuration is expected to provide a more uniform mean field for trapped electrons, and is more naturally scalable to an extended trapping region for future phases.
Data taken in this new configuration provide the new results in the following section.

\section{Latest results and conclusions}
\begin{figure}[h]
\centering
\begin{minipage}{17.5pc}
\centering
\includegraphics[width=14pc]{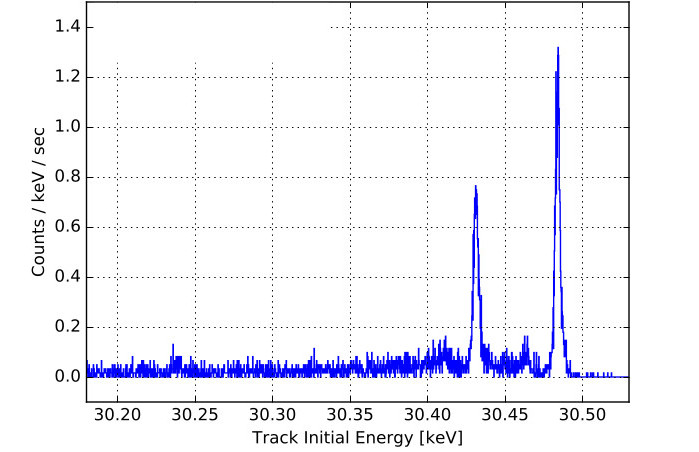}
\caption{\label{fig:spectra}Electron energy spectrum of the conversion electrons near \SI{30}{keV}. The FWHM measured in these data is less than \SI{4}{eV}, a significant improvement over the \SI{15}{eV} we have previously demonstrated.
}
\end{minipage}\hspace{1pc}
\begin{minipage}{17.5pc}
\centering
\includegraphics[width=14pc]{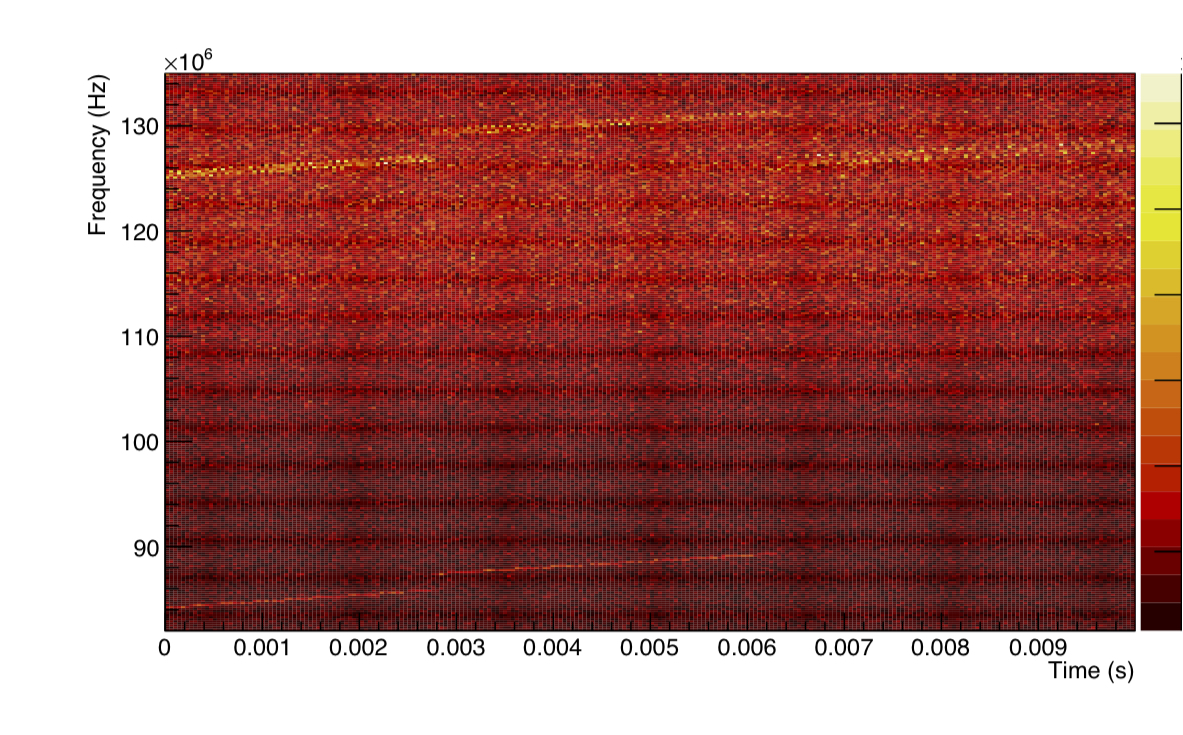}
\caption{\label{fig:sideband}Spectrogram of a single electron trapped using the two coil configuration. The cyclotron emission starts at \SI{0}{s} and \SI{85}{MHz} while the sideband begins roughly \SI{40}{MHz} higher and runs parallel to the cyclotron emission.}
\end{minipage}
\end{figure}

In the spectra from the initial phase 1 release, we demonstrated a full-width at half maximum (FWHM) of \SI{140}{eV} for the conversion electron lines near \SI{30}{keV} using a single coil field perturbation of \SI{-3.2}{mT}.
This was improved to \SI{15}{eV} when using a smaller trapping field of \SI{-1.6}{mT} at a cost of reduced trapping efficiency, and therefore event rate.
After the changes described in the previous section, data were collected using the two-coil trapping configuration, with each coil providing a \SI{3.5}{mT} perturbation.
The result for the same pair of lines near \SI{30}{keV} is shown in figure \ref{fig:spectra}, where the FWHM is measured to be less than \SI{4}{eV}.

Furthermore, trapping between two coils results in a lower axial motion frequency, bringing the sidebands produced by modulation of the cyclotron frequency within the receiver's bandwidth as shown in figure \ref{fig:sideband}.
Because sidebands were not previously apparent, they had previously not been considered in the event reconstruction software.
A significant change to that algorithm detects tracks with common start and stop times and groups them into single events, rather than falsely reconstructing the sidebands as independent electron events.
The sidebands are a result of an electron's axial motion, and detailed reconstruction of the sidebands provides an additional handle to the magnetic field experienced by confined electron.
Investigations into the use of sideband information for improved energy reconstruction are ongoing.

In conclusion, the Project 8 collaboration has previously demonstrated the feasibility of CRES for electrons energies near the \SI{18.6}{keV} endpoint of tritium.
Since then, we have achieved roughly a factor of 4 improvement in energy resolution, while moving to a more sophisticated trapping field configuration.
We have observed power modulation into sidebands and are investigating use of this additional information to improve event reconstruction.
The ongoing commissioning and operation of phase 2 will benefit from the analysis lessons learned from phase 1.

\end{document}